\documentclass[prl,preprint]{revtex4-1}
\usepackage[latin1]{inputenc}
\usepackage{amsmath}
\usepackage{amsfonts}
\usepackage{amssymb}
\usepackage{bm}
\usepackage{epsfig}
\usepackage{caption}
\usepackage{dcolumn}

\newcommand{\Eref}[1]{Eq.~(\ref{#1})}
\newcommand{\tref}[1]{Table~\ref{#1}}
\newcommand{\fref}[1]{Fig.~\ref{#1}}
\newcommand{\fermi}{G_\mathrm{F}}

\newcommand{\peven}{$\cal P$-even~}
\newcommand{\podd}{$\cal P$-odd~}
\newcommand{\sdpv}{$\cal NSD-PV$~}
\newcommand{\ka}{k_{\cal A}~}

\begin{document}

\title{Electron correlation and nuclear charge dependence of parity-violating 
properties in open-shell diatomic molecules}
\author{T.A.\ Isaev}
\email{timur.isaev@tu-darmstadt.de}
\affiliation{Clemens-Sch{\"o}pf Institute,
             TU Darmstadt, Petersenstr. 22, 64287 Darmstadt, Germany}
\author{R.\ Berger}
\email{robert.berger@tu-darmstadt.de}
\affiliation{Clemens-Sch{\"o}pf Institute,
             TU Darmstadt, Petersenstr. 22, 64287 Darmstadt, Germany}
\date{\today}
\begin{abstract}
The scaling of nuclear spin-dependent parity violating effects with
increasing nuclear charge $Z$ is discussed in two series of isovalent
open-shell diatomic molecules. The parameter $W_\mathrm{a}$ characterising
the strength of parity violation in diatomic molecules is calculated in the
framework of the zeroth-order regular approximation (ZORA) and found to be
in good agreement with the $R(Z) Z^k$ scaling law derived for
atoms in which $R(Z)$ represents a relativistic enhancement factor. The
influence of electron correlation is studied on the molecular level, with
spin-polarisation effects being conveniently accounted for by a previously
established approximate relation between the hyperfine coupling tensor and
$W_\mathrm{a}$. For high accuracy predictions of parity violating effects
in radium fluoride the necessity for systematically improvable correlation
calculations is emphasised.
\end{abstract}
\maketitle

\section{Introduction}
Molecular properties depending strongly on the behaviour of the electronic
wavefunction in the vicinity of the nuclei display a pronounced dependence
on the nuclear charge $Z$. This is known for a long time and often employed
in atomic physics to obtain qualitative estimates (see e.g.
Ref.~\cite{bouchiat:1974}). For systems with more complicated electronic
structure, in particular for molecules bearing nuclei with various $Z$
values, a simple $Z$-scaling of such molecular properties is not a priori
guaranteed, as molecular properties are also depending on the specific
nuclear arrangement, which could in principle hamper a direct comparison
(see e.g. discussion in Ref.~\cite{bakasov:2004}).  Establishing general
scaling laws also for complex molecules would present a great advantage, as
scaling laws can be used for inexpensive order-of-magnitude estimates.
Quantum chemical calculations allow to scrutinise proposed scaling laws for
a given property and investigate the form of the dependence on $Z$. This
has been applied in some detail in Ref.~\cite{berger:2008} for nuclear
spin-independent parity violating effects in chiral molecules containing
atoms from various rows of the periodic table (see
Refs.~\cite{berger:2004a,quack:2003,crassous:2005} for reviews on
molecular parity violation). In the present article we are calculating
nuclear spin-dependent parity violation interaction in alkaline earth metal
monofluorides (Mg-Ra)F (group II monofluorides) and (Zn-Cn)H (group XII
monohydrides) as an example for open-shell systems. Results and conclusions
presented below were reported by the authors on several workshops and
conferences during the years 2010 and 2011 and explicitly foreshadowed in
Ref.~\cite{Isaev:10a}. A recent paper took up the idea that we
reported on those occasions, which motivates us to present here our results
and to comment, in particular, on the inclusion of electron correlation
effects in molecular systems.

\section{Nuclear spin-dependent parity violation}
One of the properties that is predicted to depend heavily on $Z$ is the 
nuclear spin-dependent parity-odd (\podd) interaction (\sdpv), whose effective 
operator looks in a four-component (relativistic) framework like 
\cite{Moskalev:76}:
\begin{equation}
\hat{h}_\mathrm{pv}^I=\frac{\fermi}{\sqrt{2}}\sum_{A,i} k_{{\cal
A},A}\,\vec{\bm{\alpha}}\cdot \vec{I}_A\, \rho_A (\vec{r_i}),
\label{eq1}
\end{equation}
where $\fermi$ is Fermi's constant of the weak interaction, $k_{{\cal
A},A}$ is an effective parameter describing \sdpv interactions for nucleus
$A$ (caused both by the nuclear anapole moment and by weak electron-nucleon
interactions, see Ref.~\cite{flambaum:1984}), $\vec{I}_A$ and $\rho_A$ 
are the spin and nuclear spin density distribution of nucleus $A$, 
respectively. For $\vec{\bm{\alpha}}$ one uses
\begin{equation}
\vec{\bm{\alpha}}= \left(
\begin{array}{cc}
\mathbf{0} & \vec{\bm{\sigma}} \\
\vec{\bm{\sigma}} & \mathbf{0}
\end{array}
\right)
\end{equation}
with $\vec{\bm{\sigma}}$ being a vector of the $2\times2$ Pauli spin
matrices $\sigma_x$, $\sigma_y$, $\sigma_z$ and with $\mathbf{0}$ being a
$2\times2$ zero matrix. The anapole moment was proposed by Zeldovich
\cite{zeldovich:1957,zeldovich:1958} quickly after the discovery of parity
violation in processes mediated by the fundamental weak interaction. The
\emph{nuclear} anapole moment \cite{flambaum:1980} has received great
interest in atomic and molecular physics as it is caused by parity
violating interaction within the \emph{nucleus}, but should favourably be
probed in \emph{atomic} and \emph{molecular} experiments. In atoms with
stable nuclei, nuclear-spin independent terms caused by exchange of Z$^0$
bosons between nucleus and electrons typically dominate parity violating
effects and often mask those effects depending on the nuclear spin that are
significantly smaller. Thus, as of yet, only for one nucleus, namely
${}^{133}$Cs, nuclear spin-dependent parity violating effects could be
determined in atomic experiments \cite{Wood:97}. In linear open-shell
molecules, the special electronic structure itself suppresses the
contribution from nuclear spin-independent \podd terms and offers, in
principle, convenient access to nuclear-spin dependent \podd contributions
for a variety of nuclei, including those with an odd number of neutrons
instead of an odd number of protons. A complementary route would be the
detection of \sdpv in polyatomic chiral molecules by nuclear magnetic
resonance techniques
\cite{gorshkov:1982,barra:1986,barra:1988a,barra:1996,laubender:2003,soncini:2003,weijo:2005,laubender:2006,bast:2006,nahrwold:2009}.
As of yet, however, molecular parity violation has not been detected, which
underlines the particular need for identification of promising molecular
candidate systems by theoretical means.

In open-shell diatomic molecules, the contribution from interactions in
\Eref{eq1} to the effective molecular spin-rotational Hamiltonian can be
parametrised by the term $W_\mathrm{a} \ka [\vec{\lambda}{\times}\vec{S}^\mathrm{eff}]\cdot\vec{I}$ \cite{Kozlov:95}, a parity-violating contribution to the
hyperfine coupling tensor, where $\vec{\lambda}$ is the unit vector
pointing along the molecular axis, $\vec{S}^\mathrm{eff}$ is the effective
electron spin and $W_\mathrm{a}$ is a single constant characterising the $\cal
P$-odd electron spin-nuclear spin coupling for a given nucleus with nuclear
spin $\vec{I}$.  In the basis of the degenerate molecular states $\vert
\Omega \rangle $ and $\vert{-}\Omega \rangle$ ($\Omega$ is the projection of
the total electronic momentum on the molecular axis coinciding with the $z$
axis) $W_\mathrm{a}$ is approximately (see discussion below) proportional to 
the non-diagonal matrix element of the operator in \Eref{eq1}:
\begin{equation}
\label{Wa}
W_\mathrm{a}=\frac{1}{\ka [\vec{\lambda}{\times}\vec{S}^\mathrm{eff}]_{x,y}}\langle 
             \Omega \vert \frac{\partial \hat{h}_\mathrm{pv}^I}
             {\partial\vec{I}} \vert {-}\Omega \rangle_{x,y},
\end{equation}
where it was accounted for that $\vec{\lambda}$ has only a non-vanishing 
$z$-component. In contrast to $W_\mathrm{a}$, components of the hyperfine
coupling tensor $\mathbf{A}$ can be computed also as diagonal matrix
elements in the $\vert \Omega \rangle $ and $\vert {-}\Omega \rangle $ basis,
which we will exploit below to estimate spin-polarisation effects on
$W_\mathrm{a}$.

To calculate $W_\mathrm{a}$ we utilise a quasi-relativistic two-component
zeroth-order regular approximation (ZORA) approach to electroweak quantum
chemistry, which proved to perform well in calculations of the
spin-independent \podd energy differences for chiral compounds when
compared to four-component treatment \cite{
berger:2005,berger:2005a,nahrwold:2009}.  Details of the ZORA approach for
one and multielectron cases can be found elsewhere
\cite{vanLenthe:94,berger:2005,berger:2005a,berger:2008} and below we give
only the final expression of \sdpv terms in the ZORA approach in the
self-consistent field (SCF) framework of Hartree--Fock--Coulomb and
Kohn--Sham--Coulomb. The derivation of these terms together with common
consideration of the parity violation problem in open-shell polyatomic
molecules can be found in Ref.~\cite{Isaev:12}. 
\begin{table}[h!]
\tiny
\caption{Parity violating terms in ZORA Hamiltonian.}
\label{pvterms}
 \begin{tabular*}{12cm}{lll}
 Term         & Name  & Expression \\
 \hline
 $z^{(0,1)}_\mathrm{s}$       & Scalar \podd interaction &
 $\frac{\fermi}{2\sqrt{2}} Q_A \lbrace\vec{\bm{\sigma}}\cdot \vec{p},
\frac{\tilde{\omega}}{c} \rho_A(\vec{r}) \rbrace$ \\
 $z^{(1,1)}_\mathrm{hf}$      &    Scalar \podd  / hyperfine P-even interaction  &
 $\frac{\fermi}{2\sqrt{2}} Q_A \lbrace e\, \vec{\bm{\sigma}}\cdot\vec{A}_{\mu}(\vec{r}),\frac{\tilde{\omega}}{c} \rho_A (\vec{r})\rbrace$\\
 $z^{(1,1)}_\mathrm{sd}$& Nuclear spin-dependent \podd interaction &
 $\frac{\fermi}{2\sqrt{2}} 2 k_{{\cal
A},A}\lbrace\vec{\bm{\sigma}}\cdot\vec{p}, \frac{\tilde{\omega}}{c}
 \vec{\bm{\sigma}}\cdot \vec{I}_A \rho_A (\vec{r}) \rbrace$ \\
$z^{(2,1)}_\mathrm{sdr}$& Nuclear spin-dependent \podd / hyperfine P-even interaction &
 $\frac{\fermi}{2\sqrt{2}} 2 k_{{\cal A},A}
\lbrace e\, \vec{\bm{\sigma}}\cdot\vec{A}_{\mu}(\vec{r}) , \frac{\tilde{\omega}}{c}
 \vec{\bm{\sigma}}\cdot \vec{I}_A \rho_A (\vec{r}) \rbrace$ \\
\end{tabular*}
\end{table}

In \tref{pvterms}, $Q_A$ is the weak charge of nucleus $A$,
$Q_A=N_A-(1-4\sin^2\theta_\mathrm{W})Z_A$, 
where $N_A$ is the number of neutrons in nucleus $A$, $Z_A$ the nuclear
charge, $\sin^2\theta_W$ the Weinberg parameter, for which we employ the 
numerical value $\sin^2\theta_\mathrm{W}=0.2319$, and 
$\vec{A}_{\mu}$ is the magnetic vector potential from the point-like
nuclear magnetic moments $\vec{\mu}_A = \hbar \gamma_A \vec{I}_A$ with 
$\vec{A}_{\mu}(\vec{r})=(\mu_0/4\pi) \sum_A \vec{\mu}_A\times
(\vec{r}-\vec{R_A})/(|\vec{r}-\vec{R_A}|)^3$, $\gamma_A$ being the
gyromagnetic ratio and $\mu_0$ being the vacuum permeability.
The symbol $e$ denotes the elementary charge (charge of a positron),
$m_\mathrm{e}$ the mass of the electron, $\hbar = h/(2\pi)$ the reduced
Planck constant and $\lbrace x,y \rbrace = xy + yx$ the anticommutator.
The ZORA factor $\tilde{\omega}$ is also used, $\tilde{\omega}
=1/\left(2m_\mathrm{e}-{\widetilde{V}}/c^{2}\right)$, where $\widetilde{V}$ 
is the model potential (with additional damping \cite{liu:2002}) proposed 
by van W{\"u}llen \cite{wullen:1998}, which alleviates the gauge-dependence 
of ZORA. To calculate $W_\mathrm{a}$, the terms of the ZORA Hamiltonian 
which are first order in $\vec{I}$ have to be accounted for, namely
\begin{eqnarray}
\label{poddzora}
z_\mathrm{hf}^{(0,1)}+z_\mathrm{sd}^{(1,1)} = \sum_A \frac{\fermi}{2\sqrt{2}}
\left(Q_A \lbrace e\, \vec{\bm{\sigma}}\cdot\vec{A}_{\mu}(\vec{r}),\frac{\tilde{\omega}}{c} \rho_N (\vec{r})\rbrace +
2 k_{{\cal A},A} \lbrace\vec{\bm{\sigma}}\cdot\vec{p},
\frac{\tilde{\omega}}{c} \vec{\bm{\sigma}}\cdot \vec{I}_A \rho_A (\vec{r}) \rbrace
\right). 
\nonumber \\
\end{eqnarray}
An advantage of the ZORA approach is that one of the terms coupling
the \peven hyperfine interaction with the \podd nuclear spin-independent weak
interaction (the first term in \Eref{poddzora}) naturally appears after the
transition from a four-component to a two-component framework. In our
calculations we neglect this term together with accompanying response terms, 
however, as in atomic calculations it was shown to give corrections on the 
order of a few percent for heavy atoms.

\begin{figure}[h]
\begin{center}
\epsfig{file=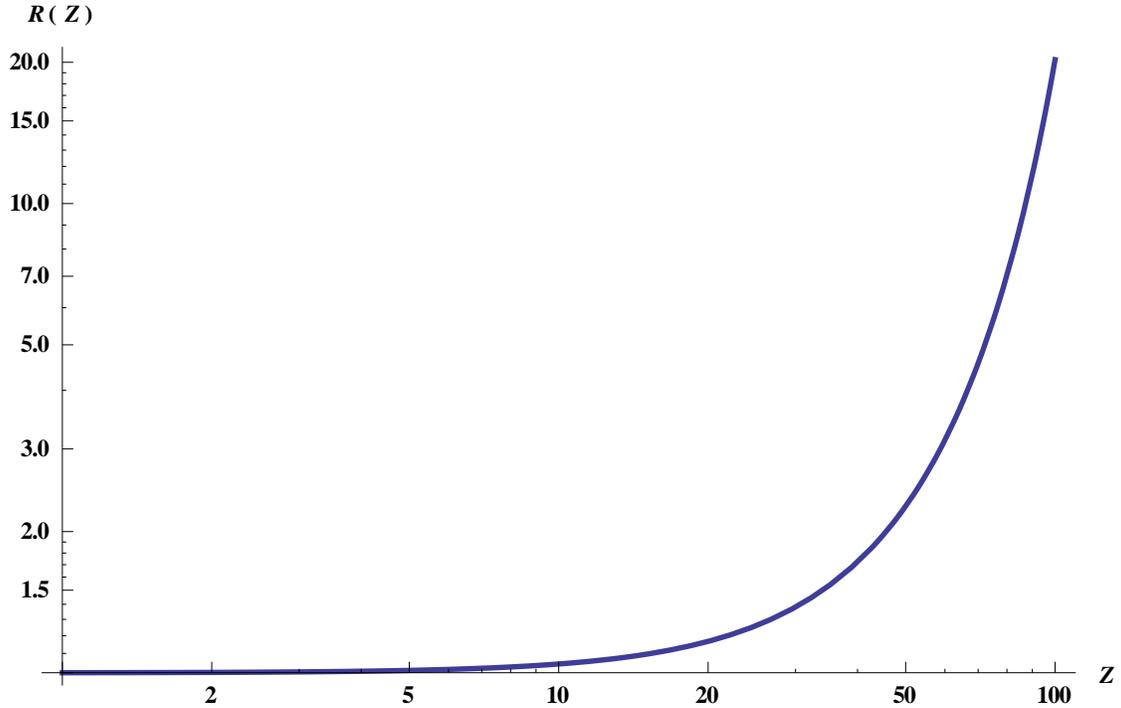,width=0.9\linewidth}
\end{center}
\caption{Relativistic enhancement factor $R(Z)$ as a function of
the nuclear charge $Z$ shown on a double logarithmic scale.\label{fig:enhancement}}
\end{figure}

The $Z$-dependent scaling behaviour of the matrix element of the nuclear
spin-independent \podd interaction was first obtained in \cite{bouchiat:1974} 
and, for nuclear spin-dependent \podd interaction, in \cite{Moskalev:76}: 
\begin{eqnarray}
\langle \mathrm{s}_{1/2}\vert\hat{h}_\mathrm{pv}^I\vert \mathrm{p}_{1/2}\rangle\sim Z^2 R(Z)\hspace*{7.5cm} \\
R(Z)=\frac{4}{3}\frac{2\sqrt{1 - 
(Z{\alpha})^2}+1}{\Gamma(2\sqrt{1 - (Z{\alpha})^2} + 1)^2}\left(\frac{a_0}{2 Z A^{1/3} \cdot r_0}\right)^{(2-2\sqrt{1 - (Z{\alpha})^2})},
\end{eqnarray} 
where $a_0$ is the Bohr radius, $\alpha$ the fine structure constant and $r_0 =
1.2~\mathrm{fm}$ a nuclear size parameter.
The analytic form of the relativistic enhancement factor $R(Z)$ was
obtained from a model treatment, such that $R(Z)$ can either 
be tabulated and used in approximate treatments or be 
calculated directly in atomic relativistic vs. non-relativistic calculations. 
As is shown in \fref{fig:enhancement} on a double logarithmic scale, $R(Z)$ 
depends heavily on $Z$ (here the analytic form was used). The proposed scaling 
behaviour for atomic systems can subsequently be studied in explicit 
calculations for molecular systems as we will show below.

\section{Calculation details and results}

The results of our study are summarised in Table \ref{all}.  We calculated
the absolute value of the parameter $W_\mathrm{a}$ for the ground
$\Sigma_{1/2}$ states of the alkaline earth metal monofluorides (Mg-Ra)F
and group XII monohydrides (Zn-Cn)H. In addition, we report results for YbF
for comparison with other approaches. In all our computations we used for
the alkaline earth metal atoms a basis set of uncontracted Gaussians with
the exponent coefficients (EC) composed as an even-tempered series. This
sequence was generated according to $\alpha_i=\gamma\beta^{N-i}$,
$i=1,\dots,N$. For $\mathrm{s}$,$\mathrm{p}$-functions $\beta$ was taken
equal to 2.0 for basis sets centred on the heavier alkaline earth metal
nuclei (Sr, Ba, Ra) as well as on ytterbium and $(5/2)^{1/25}\times10^{2/5}
\approx 2.6$ for the group XII nuclei (Zn, Cd, Hg, Cn) and the lighter
group II nuclei (Mg, Ca).  For all sets of $\mathrm{d}$-functions and
$\mathrm{f}$-functions $\beta=(5/2)^{1/25}\times10^{2/5} \approx 2.6$ was
chosen. The tighter basis sets for $\mathrm{s}$- and $\mathrm{p}$-functions
were employed, because the $\cal P$-odd operator mixes mainly $\mathrm{s}$-
and $\mathrm{p}$-waves on the heavy nucleus. For the ECs and the resulting
basis set dimensions, see \tref{basis}. On the fluorine atom in all cases
an uncontracted ANO basis set of triple-zeta quality \cite{emsl} and on
hydrogen an {$\mathrm{s}$,$\mathrm{p}$}-subset of an uncontracted
correlation-consistent basis set of quadruple-zeta quality \cite{emsl} were
used with the ECs given explicitly in \tref{basis}. 

The nuclear density was modelled by a spherical Gaussian distribution $\rho
(R)=\rho_0 e^{-\frac{3}{2\xi}R^2}$, where $\xi$ is the mean root square
radius of the corresponding nucleus computed according to the empirical
formula $\xi=(0.836 A^{1/3}+0.57)~\mathrm{fm} = (1.5798
A^{1/3}+1.077)~10^{-5}a_0$, where $A$ is the given mass number of the
respective isotope. Within this work we employed mass numbers corresponding
to the standard relative atomic mass rounded to the nearest integer, namely
${}^{1}$H, ${}^{19}$F, ${}^{24}$Mg, ${}^{40}$Ca, ${}^{65}$Zn, ${}^{88}$Sr,
${}^{112}$Cd, ${}^{137}$Ba and ${}^{201}$Hg.  The radium nucleus with
atomic mass number ${225}$ was taken and the copernicium nucleus with
atomic mass number ${284}$.  As computed $W_\mathrm{a}$ values do not
depend too strongly on the atomic mass number (for ${}^{213}$RaF,
${}^{223}$RaF and ${}^{225}$RaF changes in $|W_\mathrm{a}|$ were found on
the order of a few Hz \cite{Isaev:10a}), we report only one value for
$|W_\mathrm{a}|$, even though the specific isotope corresponding to the
standard relative atomic mass may have a nuclear spin quantum number of
$I=0$.  A modified version
\cite{berger:2005,berger:2005a,nahrwold:2009,Isaev:12} of the program
package {\sc TURBOMOLE} \cite{Alrichs:89} was used for the complex
generalised SCF (Hartree--Fock or Kohn--Sham) calculations. As spatial
symmetry was not exploited, we also calculated the value of the projection
$\Omega$ of the total electron angular momentum on the molecular axis. In
the two-component generalised Hartree--Fock (GHF) calculations
$\Omega=0.5\pm 10^{-3}$, in the two-component density functional theory
(DFT) calculations within the generalised Kohn--Sham (GKS) framework
$\Omega=0.5\pm 10^{-4}$.  The value of $|W_\mathrm{a}|$ was calculated
according to \Eref{Wa}. In the ZORA calculations $\vert\Omega\rangle$ and
$\vert-\Omega\rangle$ are many-electron functions and the L{\"o}wdin
formula \cite{Lowdin:55} for calculations of the matrix elements between
non-orthogonal one-determinantal (OD) functions was applied:
\begin{eqnarray}
 \langle\Psi_1\vert \widehat{W} \vert \Psi_2 \rangle=\sum_k\sum_l\langle\tilde{\psi}_k\vert \widehat{w} \vert \psi_l \rangle
{\cal D}(k \vert l),
\end{eqnarray}
where $\vert \Psi_1 \rangle $ and $\vert \Psi_2 \rangle $ can be either
orthonormalised or non-orthonormalised OD functions,
$\langle\tilde{\psi}_l\vert \widehat{w} \vert \psi_k \rangle$ is the matrix
element of the one-electron operator $\widehat{w}$ between members of the
two sets of molecular spin-orbitals with
$\langle\tilde{\psi}_i|\tilde{\psi}_j\rangle=\delta_{ij},
\langle\psi_i|\psi_j\rangle=\delta_{ij}$, that are occupied in the OD
wavefunctions $\Psi_1$ and $\Psi_2$, respectively, and ${\cal D}(k \vert
l)$ is the minor of $\mathbf{S}$ of rank $n-1$ ($n$ is the number of
electrons) which is obtained from the original OD wavefunctions by crossing
out in the overlap matrix $\mathbf{S}$ (that has the matrix elements
$s_{kl} = \langle\tilde{\psi}_k|\psi_l\rangle$) the $k$-th row and $l$-th
column with subsequently forming the determinant of the resulting
submatrix.  We note in passing that in the direct application of the
present complex GHF/GKS approach only the absolute value of $W_\mathrm{a}$
is immediately accessible, whereas determination of its sign requires an
additional symmetrisation procedure, which is for the purpose of the
present study, however, not required.

In calculations of the (Mg-Ra)F row two different exchange-correlation (XC) 
functionals were used in a generalised Kohn--Sham DFT framework: 1) the 
local density approximation {\sc LDA} and 2) the Becke 3-parameter hybrid 
functional containing Becke's exchange functional together with
Lee-Yang-Parr's (LYP) correlation functional {\sc B3LYP}. 
This latter hybrid XC functional, which contains an admixture
of about 20~\% non-local Fock exchange, was used in the form employed in the 
{\sc Gaussian 03} program package \cite{g03} with an approximation
(VWN3) to the correlation functional of the homogeneous electron gas.
The equilibrium distance for all diatomic molecules was taken from 
experimental data, except for RaF, where the distance was obtained in
\cite{Isaev:10a} from four-component relativistic coupled cluster calculations 
in the Fock space (FS-RCC-SD) and CnH, where we used the bond length
obtained in the two-component GHF framework.

\begin{figure}[b]
\begin{center}
\epsfig{file=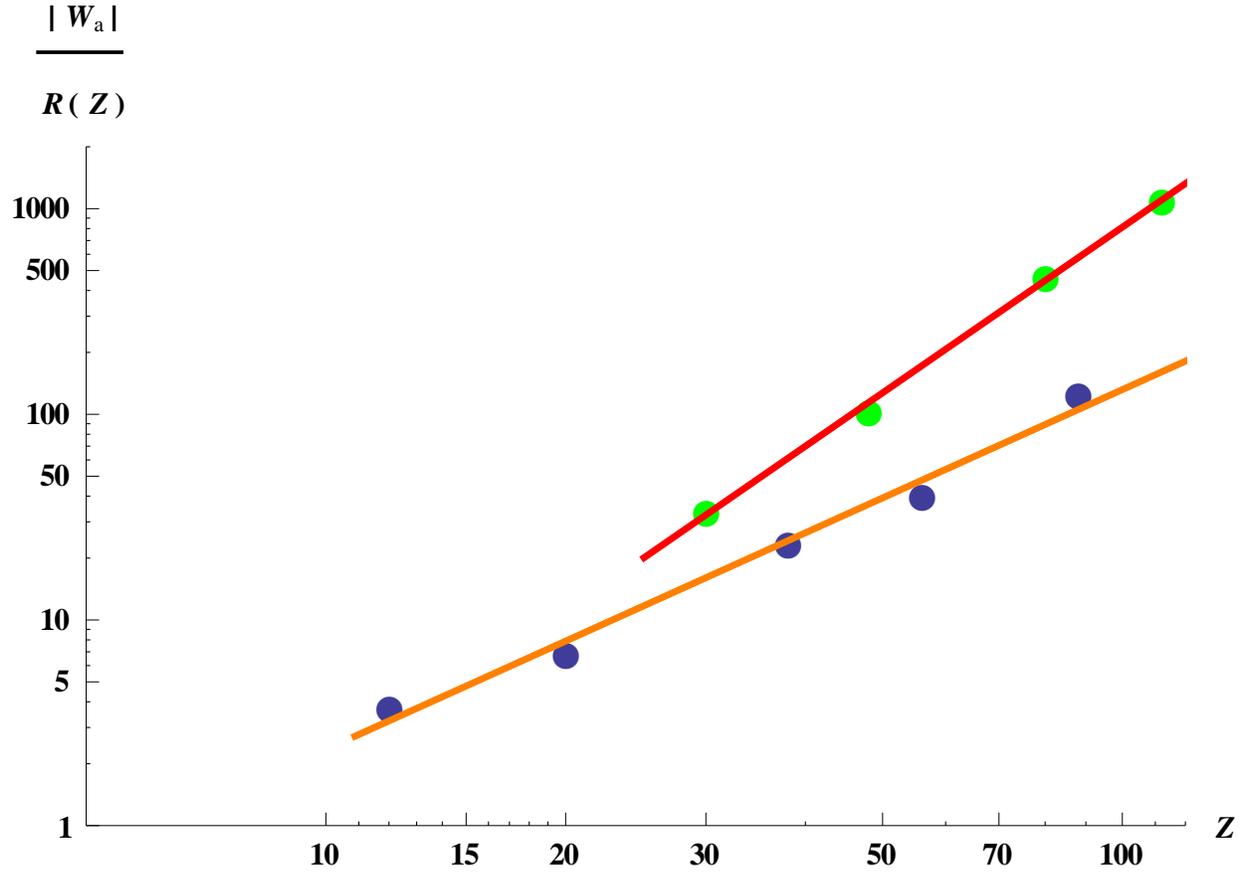,width=\linewidth}
\end{center}
\caption{Scaling of $|W_\mathrm{a}/R(Z)|$ GHF values (in Hz) with $Z$ for the
(Mg-Ra)F and (Zn-Cn)H series (orange line, blue dots and red line, green
dots, respectively) on a double logarithmic scale. The slope of
the lines is 1.75 for (Mg-Ra)F and 2.68 for (Zn-Cn)H, which implies
a $R(Z) Z^k$ scaling law for $W_\mathrm{a}$ with $k=1.75$ and $k=2.68$,
respectively.\label{fig:results}}
\end{figure}

One can see from \tref{pvterms} two main trends when accounting for
correlations by DFT: 1) systematic increase in the value of $|W_\mathrm{a}|$
from {\sc B3LYP} to {\sc LDA} XC functionals and 2) relative decrease in
correlation contributions from 33~\% for MgF to about 12~\% for RaF. Both
these dependences are consistent with previous observations and
anticipations. The former trend was observed for parity violating energy
differences between enantiomers of chiral molecules \cite{berger:2005a},
the latter is also not surprising as the main contribution in this class of
heavy-atom open-shell diatomic molecules is expected to arise from
spin-polarisation effects, which cannot (fully) be accounted for by direct
calculation of non-diagonal matrix elements between complex GHF
wavefunctions at least for $\cal T$-odd operators, for which thus results
of essentially paired GHF quality are obtained. A discussion of the
influence of symmetry breaking for OD wavefunctions on matrix elements of
different operators can be found in Ref.~\cite{Isaev:12b}.  Finally we plotted
on a double logarithmic scale (\fref{fig:results}) instead of
$|W_\mathrm{a}|$ the values of $|W_\mathrm{a}/R(Z)|$ obtained on the GHF level
against $Z$, as we have argued previously \cite{berger:2008,Isaev:10a} that
one should correct for the relativistic enhancement factors when attempting
to extract $Z^k$ scaling laws from quasi-relativistic and relativistic
calculations. Fitting of the points in \fref{fig:results} by a linear
function gives a slope equal 1.75 for (Mg-Ra)F and 2.68 for (Zn-Cn)H which
is indeed close to the scaling factor for \sdpv interaction matrix element.

Our current results have been mainly confirmed by recent four-component
calculations of $W_\mathrm{a}$ in the series of diatomic radicals
(Mg-Ra)F~\cite{Borschevsky:12}. The authors of \cite{Borschevsky:12} have
also observed the $Z^k$ scaling (also with $k$ close to 2) for
$W_\mathrm{a}/R(Z)$.  Besides performing Dirac--Hartree--Fock--Coulomb and
Dirac--Kohn--Sham calculations in a paired GHF and paired GKS framework,
which can not account for core-polarisation effects, the authors of
Ref.~\cite{Borschevsky:12} employed some approximate atom based schemes to
roughly estimate part of electron correlation effects via scaling factors.
Our treatment, however, is based on the complex GHF/GKS framework and thus
allows to capture part of the electron correlation effects directly within
the molecular calculations (see also below), whereas some contributions are
not included due to calculation via off-diagonal matrix elements between
time-reversed wavefunctions.  Our direct DFT-based estimates for
$|W_\mathrm{a}|$ in MgF, CaF and SrF can reasonably well be reproduced by
the indirect procedure employed in Ref.~\cite{Borschevsky:12}. For BaF and
RaF we find, however, only a modest electron correlation contribution on
the DFT level of theory whereas in Ref.~\cite{Borschevsky:12} significant
changes are reported for RaF. Even without the subsequent attempts to
account for further electron correlation effects, in
Ref.~\cite{Borschevsky:12} DHF and DKS values for $|W_\mathrm{a}|$ in RaF
differ by more than 15~\% (and by about 30~\% for LDA XC functional) and in
BaF only by 2~\%.  As it was mentioned, the latter result (for BaF) is
actually in agreement with the earlier calculations in
Ref.~\cite{Kozlov:97}, in which the authors found that the main
contribution comes from the spin-polarisation effects, though further
accounting for electron correlation gives minor contribution. We note in
passing that in Ref.~\cite{Borschevsky:12} the data are mixed up for the
calculations without accounting for spin-polarisation (SCF in notations of
the authors of Ref.~\cite{Kozlov:97}, $W_\mathrm{a}$ = 111~Hz), with the
accounting for spin-polarisation (SCF-EO, $W_\mathrm{a}$ = 181~Hz)  and
electron correlation + spin-polarisation (RASSCF-EO, $W_\mathrm{a}$ =
175~Hz). In Ref.~\cite{Borschevsky:12} Faegri's energy-optimised basis sets
were employed, which required augmenting with additional functions to be
used in calculations of properties that depend on the behaviour of the
electronic wavefunction near the nucleus. To check the influence of the
basis set choice, we performed calculations with two additional basis sets
(see below for basis set specification) for Ra together with an
uncontracted aug-cc-pVTZ basis set 11s6p3d2f on fluorine nucleus. The first
Ra basis set (Basis S in \tref{all}) was Faegri's uncontracted basis set
25s21p14d9f recommended for relativistic calculations
\cite{Faegri:internet} and another one (Basis L) was a large even-tempered
basis set 36s33p22d15f generated according to recommendation of the article
\cite{Faegri:05}.  The result of the calculations with these basis sets
clearly shows that with the extension of the basis set from Basis S to
Basis L the difference between GHF and GKS results of essentially paired
generalised SCF quality decreases from 38~\% to 20~\% for LDA XC
functional, getting close to the values reported by us ($\approx 12~\%$)
for the basis sets we used herein and in Ref.~\cite{Isaev:10a}.  This
provides some indication that the pronounced electron correlation effects
reported in Ref.~\cite{Borschevsky:12} for the DFT framework might
primarily be caused by the special choice of basis set therein.

To estimate spin-polarisation contributions within the GHF approach we use
scaling relations from the semiempirical molecular model by
Kozlov~\cite{Kozlov:85}, which is known to reproduce {\it ab initio}
parameters of the $\mathcal{P},\mathcal{T}$-odd spin-rotational Hamiltonian
for ground states of BaF, YbF and some other molecules with an accuracy of
10-15\%. For this model some simple (approximate) relations can be
established between the parameters of the electronic structure, required
for calculations of $W_\mathrm{a}$ of the linear diatomic molecules
employed in the current work in their $\Sigma_{1/2}$ ground states, and the
hyperfine coupling tensor terms $A_\mathrm{iso}$ (isotropic) and
$A_\mathrm{d}$ (dipole).  For our purpose and the current set of molecules,
however, more important is that the relation between $W_\mathrm{a}$
obtained for different approximations (e.g.  complex and paired generalised
Hartree--Fock wavefunctions, cGHF and pGHF, respectively) is approximately
equal to the ratio between the square root of the products of
$A_\mathrm{iso}$ and $A_\mathrm{d}$ (see equations (33),(34) and (10) in
Ref.~\cite{Kozlov:85}; assuming that signs of $A_\mathrm{iso}$,
$A_\mathrm{d}$ are identical):
\begin{equation}
\frac{W_\mathrm{a}^\mathrm{cGHF}}{W_\mathrm{a}^\mathrm{pGHF}}\approx
\left[\frac{(A_\mathrm{iso}\cdot A_\mathrm{d})^\mathrm{cGHF}}
           {(A_\mathrm{iso}\cdot A_\mathrm{d})^\mathrm{pGHF}}\right]^{1/2}.
\end{equation}
Thus, by calculating the hyperfine tensors with accounting for
spin-polarisation (in our case as diagonal matrix elements within the
complex GHF scheme) and without accounting for it (as non-diagonal matrix
elements, leading to results of essentially paired GHF quality) we can
restore spin-polarisation contributions, which are expected to be most
important for RaF. The results of this scaling are presented in
\tref{scale}. One can observe that for molecules with a valence electronic
structure similar to RaF the relative deviation of our scaling for
$W_\mathrm{a}$ parameters is better than 10~\% when judged from the
corresponding RASSCF/EO results or about 10~\% in comparison with the
semiempirical estimates for HgH by Kozlov~\cite{Kozlov:85}. This finding is
particularly encouraging for the identification of promising molecular
candidates, although we expect the accuracy in general to be somewhat lower
than implied by the present results.  Thus, for reliable estimates of
electron correlation and spin-polarisation effects on the value of
$W_\mathrm{a}$ (and other properties depending on the behaviour of the
wavefunction near the nucleus) in RaF one has to employ high-order
correlation calculations, for instance similar to those in
Ref.~\cite{Isaev:04}. It is also interesting to note that in the group XII
monohydrides series the spin-polarisation contribution should
suppress rather than enhance the \sdpv interaction, at least for the two heavy
representatives reported in \tref{scale}.

Finally, we emphasise that although the treatment of relativistic effects
in the four- and two-component framework is different, deviations between 
results for \sdpv operators are not expected to be significantly larger than 
3~\% for the heavier nuclei (row 4--7) when judged on the basis of earlier 
calculations \cite{berger:2005a,nahrwold:2009}, provided appropriate basis 
sets are used (see also discussion in Ref.~\cite{berger:2008}).

\begin{table*}[b]
\tiny
\caption
 {Basis sets parameters for ZORA HF/DFT  calculations.
 Even-tempered basis sets of uncontracted Gaussians are given in the form 
 $N_\mathrm{bas}$; $l$; (EC$_\mathrm{max}$; EC$_\mathrm{min}$), where
 $N_\mathrm{bas}$ is the number of Gaussians, $l$ is 
 $\mathrm{s}$, $\mathrm{p}$, $\mathrm{d}$ or $\mathrm{f}$ 
 and represents the angular momentum quantum numbers 0, 1, 2 or 3. 
 EC$_\mathrm{max}$ and EC$_\mathrm{min}$ are the largest and smallest
 exponent coefficients, respectively.
}
\begin{tabular*}{\linewidth}{rrlrlrlrl}
  & \multicolumn{2}{c}{Mg}& \multicolumn{2}{c}{Ca}&\multicolumn{2}{c}{Sr,Ba,Yb}& \multicolumn{2}{c}{Ra}\\ \hline
  & 27; s; & (500000000; 0.00769) & 27; s; & (500000000; 0.00769) &  37; s; & (2000000000; 0.0291)   & 39; s; & (2000000000; 0.00728)  \\
  & 25; p; & (191890027; 0.02000) & 25; p; & (191890027; 0.0200)  &  34; p; & (~500000000; 0.0582)   & 34; p; & (~500000000; 0.0582)  \\
  &  4; d; & (3.750; 0.21336)     & 13; d; & (13300.758; 0.135789)&  14; d; & (13300.758; 0.0521)    & 14; d; & (13300.758; 0.0521) \\
  &        &                      &        &                      &   9; f; & (751.8368350; 0.3546)  & 10; f; & (751.8368350; 0.13638) \\
\hline
\end{tabular*}

\begin{tabular*}{\linewidth}{rrl}
       & \multicolumn{2}{c}{Zn,Cd,Hg,Cn} \\
       & 27; s; & (500000000; 0.0077)       \\
       & 25; p; & (191890027; 0.0200)       \\
       & 14; d; & (13300.758; 0.0521)       \\
       &  8; f; & (751.8368350; 0.9219352)  \\
\hline
\end{tabular*}

\newcolumntype{d}{D{.}{.}{-5}}
\begin{tabular*}{\linewidth}{ldddldd}
& \multicolumn{3}{c}{F ANO basis} & &\multicolumn{2}{c}{H basis} \\
& \multicolumn{1}{c}{$\mathrm{s}$} 
   & \multicolumn{1}{c}{$\mathrm{p}$} 
      & \multicolumn{1}{c}{$\mathrm{d}$}                        
         & ~~~& \multicolumn{1}{c}{$\mathrm{s}$} 
                & \multicolumn{1}{c}{$\mathrm{p}$} \\
 & 103109.46  &  245.33029    & 5.000000 & & 82.640 & 2.2920000 \\ 
 & 15281.007  &  56.919005    & 1.750000 & & 12.410 & 0.8380000\\
 & 3441.5392  &  17.604568    & 0.612500 & & 2.8240 & 0.2920000\\
 & 967.09483  &  6.2749950    & 0.214375 & & 0.7977 &\\
 & 314.03534  &  2.4470300    &          & & 0.2581 &\\
 & 113.44230  &  0.9950600    &          & & 0.08989 &\\
 & 44.644727  &  0.4039730    &          & & & \\
 & 18.942874  &  0.1548100    &          & & & \\
 & 8.5327430  &  0.0541840    &          & & & \\
 & 3.9194010  &               &          & & & \\
 & 1.5681570  &               &          & & & \\
 & 0.6232900  &               &          & & & \\
 & 0.2408610  &               &          & & & \\
 & 0.0843010  &               &          & & & \\

\hline
\end{tabular*}

\label{basis}
\end{table*}


\begin{table*}
\caption
 {Calculated $\cal P$-odd parameter $|W_\mathrm{a}|$ (in Hz) for open-shell
  diatomic molecules together with the charge number $Z$ of the heavy nucleus
  and the equilibrium distance $R_\mathrm{e}$ employed. 
}
\begin{tabular*}{\linewidth}{lrdddd}
        &     &  & \multicolumn{3}{c}{$|W_\mathrm{a}|$/Hz}\\\cline{4-6}
        & $Z$ &  \multicolumn{1}{c}{$R_\mathrm{e}/a_0$} 
                 & \multicolumn{1}{c}{GHF} 
                   & \multicolumn{1}{c}{GKS/B3LYP} 
                     & \multicolumn{1}{c}{GKS/LDA} \\
\hline
MgF     &  12 &    3.30    &  3.9   &    4.9    &   5.2       \\
CaF     &  20 &    3.71    &  8.0   &    9.2    &   9.5       \\
SrF     &  38 &    3.92    &  3.9~\times 10^{1}   &    4.6~\times 10^{1}     &    4.8~\times 10^{1}      \\
BaF     &  56 &    4.07    &  1.11\times 10^{2}{\,}^\mathrm{a}   &   1.19\times 10^{2}     &   1.25\times 10^{2}      \\ 
RaF     &  88 &    4.24    &  1.30\times 10^{3}{\,}^\mathrm{b}   &  1.42\times 10^{3}     &  1.47\times 10^{3}      \\ \hline
RaF (Basis S) &  88 &    4.24    & 1.07\times 10^{3}   &       &  1.48\times 10^{3}      \\
RaF (Basis L) &  88 &    4.24    & 1.24\times 10^{3}   &       &  1.50\times 10^{3}      \\\hline
ZnH     &  30 &    3.01    &  4.7~\times 10^{1}   &         &          \\
CdH     &  48 &    3.36    &  2.23\times 10^{2}   &        &         \\ 
HgH     &  80 &    3.33    &  3.30\times 10^{3}{\,}^\mathrm{c}   &       &        \\
CnH     & 112 &    3.10    &  4.88\times 10^{4}               &       &        \\\hline
\multicolumn{6}{l}{\footnotesize a) In Ref.~\cite{Isaev:10a} 111 Hz were
obtained with a slightly different basis set.} \\
\multicolumn{6}{l}{\footnotesize b) In Ref.~\cite{Isaev:10a} 1.3~kHz were
reported for a slightly different basis set.}\\
\multicolumn{6}{l}{\footnotesize c) In Ref.~\cite{Isaev:10a} 3.3~kHz were
reported for a slightly smaller basis set.}\\
\end{tabular*}
\\
\label{all}
\vspace{0.5cm}
\end{table*}


\begin{table*}
\caption
 {Calculated and scaled {\it ab initio} values for the parameter
$|W_\mathrm{a}|$ in BaF, YbF and RaF together with scaling factor
$f=\left[\frac{(A_\mathrm{iso}\cdot A_\mathrm{d})^\mathrm{cGHF}}
              {(A_\mathrm{iso}\cdot A_\mathrm{d})^\mathrm{pGHF}}\right]^{1/2}$. 
 Additional {\it ab initio} results and calculation methods are taken from
 the corresponding references. A bond length of $3.80~a_0$ was used in the
 calculation for YbF.
}
\begin{tabular*}{\linewidth}{lddddcr}
 & \multicolumn{3}{c}{$|W_\mathrm{a}|$/Hz} & & Method & Ref.\\
\cline{2-4}
        &   \multicolumn{1}{c}{GHF} 
                             &  \multicolumn{1}{c}{Scaled} 
                                                 & \multicolumn{1}{c}{\emph{ab initio}} &f & &\\
\hline
BaF     & 1.11\times 10^{2}  &  1.9\times 10^{2} &  1.81\times 10^{2}  &  1.68  & ~~~~~~SCF/EO    & \cite{Kozlov:97}   \\ 
        &       &      &  1.75\times 10^{2}  &        & ~RASSCF/EO & \cite{Kozlov:97}      \\
YbF     & 4.65\times 10^{2}  & 6.1\times 10^{2} &  6.34\times 10^{2}  &  1.31  & ~RASSCF/EO &\cite{Mosyagin:98}  \\ 
RaF     & 1.30\times 10^{3}  & 2.1\times 10^{3} &       &  1.65  &            &     \\
HgH     & 3.30\times 10^{3}  & 2.0\times 10^{3} &  \mathrm{a}     &  0.62  &            &     \\
CnH     & 4.88\times 10^{4}  & 3.1\times 10^{4} &       &  0.63  &            &     \\\hline
\multicolumn{7}{l}{\footnotesize a) Semi-empirical estimate of
Ref.~\cite{Kozlov:85} based on spectroscopic parameters of ${}^{199}$HgH
and ${}^{201}$HgH:} \\
\multicolumn{7}{l}{\footnotesize \phantom{a)}
$W_\mathrm{a}=1800~\mathrm{Hz}$ and
$W_\mathrm{a}=1940~\mathrm{Hz}$, respectively.}
\end{tabular*}
\label{scale}
\end{table*}

\clearpage
\section{Conclusions}
We have reported herein a numerical study on nuclear charge dependent
scaling of molecular properties in open-shell diatomic molecules. After we
have accounted for a relativistic enhancement factor $R(Z)$, which grows
non-polynominally with the nuclear charge $Z$, we obtain an approximate
$Z^2$ scaling behaviour for the nuclear spin-dependent parity violating
parameter $W_\mathrm{a}$ computed at the respective equilibrium structures.
This term contributes to the effective spin-rotational Hamiltonian used for
high-resolution studies which aim for the first detection of molecular
parity violation. The present confirmation of a simple scaling law is
excellent news as it allows for quick estimates of parity violating effects
in a whole series of diatomic molecules. Within the complex generalised
Kohn--Sham framework employed in this work, part of electron correlation
effects on this molecular property can be accounted for, although some
contributions are still missing. Spin-polarisation effects can
approximately be included within a simple, but powerful molecular scaling
scheme utilised previously for semi-empirical estimates. For high accuracy
calculations, more sophisticated molecular electron correlation approaches
are clearly needed, but given the present experimental status, the current
approximate approaches allow to identify promising molecular candidates
such as RaF, which was proposed in Ref.~\cite{Isaev:10a}.
\section{Acknowledgement}
We are indebted to Mikhail Kozlov and Sophie Nahrwold for discussions and
are particularly thankful for numerous discussions at the 2010 ECT*
workshop on "Violations of discrete symmetries in atoms and nuclei" in
Trento. Financial support by the Volkswagen Foundation and computer time
provided by the Center for Scientific Computing (CSC) Frankfurt are
gratefully acknowledged.

\end{document}